\documentclass[conference]{IEEEtran}
\ifCLASSINFOpdf
\else
\fi
\usepackage{pgfplots}
\usepackage{comment}
\usepackage{breqn}
\usepackage{booktabs}
\usepackage{balance}

\usepackage{hyperref}
\makeatletter
\newcommand{\printfnsymbol}[1]{%
  \textsuperscript{\@fnsymbol{#1}}%
}
\makeatother
\IEEEoverridecommandlockouts
\begin{document}
%
\title{A Survey of Novel Cache Hierarchy Designs for High Workloads}

\author{
\IEEEauthorblockN{Pranjal Rajput\printfnsymbol{1}}
\IEEEauthorblockA{
Delft University of Technology}
\and

\IEEEauthorblockN{Sonnya Dellarosa\printfnsymbol{1}\thanks{\printfnsymbol{1}Denotes equal contribution}}
\IEEEauthorblockA{
Delft University of Technology}
\and

\IEEEauthorblockN{Kanya Satis\printfnsymbol{1}}
\IEEEauthorblockA{
Delft University of Technology}

%
}


\maketitle

\begin{abstract}
Traditional on-die, three-level cache hierarchy design is very commonly used but is also prone to latency, especially at the Level 2 (L2) cache. We discuss three distinct ways of improving this design in order to have better performance. Performance is especially important for systems with high workloads. The first method proposes to eliminate L2 altogether while proposing a new prefetching technique, the second method suggests increasing the size of L2, while the last method advocates the implementation of optical caches. After carefully contemplating results in performance gains and the advantages and disadvantages of each method, we found the last method to be the best of the three.
\end{abstract}

\IEEEpeerreviewmaketitle

\section{Introduction}


In the world of server computing, one of the main aspects that is constantly being improved is speed. While there are many things that contribute to the speed of a server computer, cache plays a fairly crucial role in being able to increase it. This is the reason why we chose cache hierarchy design as the topic of our assignment. 

The main objective of this paper is to discuss three different approaches of cache hierarchy design that could improve the performance of a system with high workloads in comparison to a traditional three-level hierarchy design. The there levels are called Level 1 cache (L1), Level 2 cache (L2), and Last Level Cache (LLC). L1 has the least amount of memory latency. Cache is also mostly on die, which means that unlike memory, cache is usually fabricated on the same chip as the processor.

This survey will look into three different perspectives, offered by three different studies~\cite{paper1}~\cite{paper2}~\cite{paper3}, in tackling this issue. One solution suggests a new algorithm in servicing cache requests, one examines different sizes of the physical architecture of the hardware, while one investigates an optical bus-based architecture.


After a careful examination of the three possible solutions by looking at its benefits, drawbacks, and most importantly, the performance gains, we conclude that Solution 3 is the most optimal solution. The reason for this is that Solution 3 manages to considerably improve the performance for high workloads while also addressing the problems that arise in the other two solutions.

In Section II, we provide a detailed description of each solution in separate sub-sections. In Section III, we compare the three different solutions, in terms of their performance gains and the benefits and consequences of each of them. We conclude the survey in Section IV.

\section{Description of the Evaluated Solutions}

The first approach, described in Section II-A, considers the criticality of instructions, where those of higher ones should be served as quickly as possible, i.e. by the L1 cache. We will define the meaning of criticality, specify a particular way to describe it, and introduce a new pre-fetching technique that can be used to ensure that the critical instructions are served by the L1 cache. The second approach, discussed in Section II-B, discusses increasing the L2 cache size so as to increase the L2 hit rates and thus decrease the shared cache access latency. They also discuss the implementation of an exclusive hierarchy to stay in accordance with the design policies and how this change increases performance as well. Lastly, the third approach, outlined in Section II-C, is about implementing optical caches and how they are a significant improvement to the conventional cache hierarchies.

\subsection{Solution 1}


A conventional cache hierarchy design typically implements an egalitarian approach, where all instructions are treated as equals and none has a higher priority than others. This is actually not ideal, some instructions may be more damaging to the performance than others, as these instructions could cost more number of cycles. A lack of focus in such bottleneck-inducing issue forms the fundamental motivation for \textbf{Criticality Aware Tiered Cache Hierarchy (CATCH)}, proposed by Nori et al. ~\cite{paper1} as a possible solution.

The main goal of CATCH is to first identify the criticality of the program at the hardware level and then use new inter-cache prefetching techniques such that data accesses that are on a \textit{critical} execution path are serviced by the innermost L1 cache, instead of the outer levels L2 or LLC.

Criticality in CATCH is calculated with the help of the data dependency graph (DDG) of the program. In this graph, which was first introduced by Fields et al.~\cite{fields2001} each instruction has three nodes. One node denotes the allocation in the core, another node denotes the dispatch to execution units and the last node denotes the write-back time. The links between nodes of two different instructions represent the dependencies between the two. After an instruction completes, its nodes are created in the graph. After the instruction has been buffered at least two times the reorder buffer size, they are able to identify the critical path of this particular instruction, by finding the path with the most weighted length. They then walk through this critical path and mark the critical load instruction addresses, otherwise known as Program Counters (PC), that were on the path. For a processor with a reorder buffer size of 224, the total area required for this critical path identification on the hardware is about 3 KB. The area calculation is discussed in details in~\cite{paper1}.


However, it is important to note that criticality is not only affected by the executed application, but also by the hardware, due to its responsibility in determining the execution latency of any instruction. This means that the critical paths will dynamically change. It is therefore important to dynamically calculate the critical paths throughout the execution of the application.

Now that they have identified the critical loads, they need to prefetch these loads in the L1 cache. As such, the cache lines that exist in the outer cache levels and correspond to the critical load accesses need to be prefetched into L1 in a timely manner. Due to its size, L1 has a small capacity and bandwidth. It is therefore imperative to only prefetch a select set of critical loads that affect the performance the most. Additionally, they also need to avoid overfetching into the L1 cache as this might cause L1 thrashing, where new critical paths may be formed and performance will hinder performance. Thus, the authors propose Timeliness Aware and Criticality Triggered prefetchers (TACT), which prefetch the data cache lines from L2 and LLC into the L1 cache just in time before they are needed by the core.

Cache prefetching is a function of \textit{Target-PC}, which is the load PC that needs to be prefetched, \textit{Trigger-PC}, which is the load instruction that triggers the prefetching of the Target, and \textit{Association}, which is the relation between the attributes of the Trigger-PC and the address of the Target-PC. The goal of TACT is to learn this relation to allow timely prefetching for the Target-PC. To this end, TACT is split into three components, TACT Cross, TACT Deep Self, and TACT Feeder.

Cross associations of trigger address often appear when Trigger-PC and Target-PC of a load instruction have the same base register but different offsets. TACT Cross involves a mechanism in the hardware that can spot these cross associations in order to utilize them. First, they note that more than 85\% delta values of cross address association are observed to be within a 4 KB page, which indicates a very high chance that the Trigger-PC accesses the same 4 KB page as the Target-PC. They then monitor the last 64 4 KB pages, 
until one the Target-PC has correctly identified one stable Trigger-PC. After this training, TACT Cross will be able to provide a prefetch just in time before the core dispatches the Trigger-PC.

TACT employs a stride prefetching method, though the typically used prefetch distance is actually not punctual enough to hide all the L2 and LLC hit latency. On the other hand, if the prefetch distance of all load PCs is increased, there would be too many prefetches that may worsen the L1 latency and negatively affect performance. TACT Deep Self, therefore, increases the prefetch distance of a select critical load PCs in order to avoid hurting performance.

The last component, TACT Feeder, is necessary for when there exist no address associations for the critical loads, in which case TACT will attempt data associations. TACT Feeders will track the dependencies between load instructions and establish a linear function of the form $Address = Scale\times Data+Base$ between the Target PC address and the Trigger data until stables values for the variables $Scale$ and $Base$ are identified. The data from the Trigger-PC can then prompt a prefetch for the Target-PC.

They performed a preliminary investigation to examine the performance gain of such prefetching technique and found that a configuration with and without L2, for high amounts of PCs, yield similar performance gains. Supported by the authors' recognition of the current trend of increasing the size of L2, the benefits of which they identified as caused mainly by the critical loads hitting the L2 cache, stems the idea of eliminating L2 from the architecture of the cache hierarchy. Thus, with CATCH, L1 serves the primary purpose of servicing the critical loads, while LLC functions in reducing memory misses and preventing the formation of new critical paths. A two-level cache hierarchy would consequentially reduce the chip area redundancies and provide more room to increase the size of LLC or new cores. Although, they note that this significantly increases the interconnect traffic. The total area necessary for all the TACT components is around 1.2 KB.

For evaluation, benchmarks of single thread workloads are executed on a configuration with a large L2 (1 MB) and a shared, exclusive LLC of different sizes. The performance gain or loss are examined after the L2 is removed and then workloads are re-tested without then with CATCH. There seems to be a performance gain of 10\% on a server. With CATCH, a higher number of workloads and a bigger LLC size result in better performance gains. CATCH is also tested for multi-programmed workloads, where a three-level CATCH-implemented hierarchy gives better performance by 8.95\%, while a two-level CATCH-implemented hierarchy yields a performance gain of 8.45\%.

\subsection{Solution 2}

Another solution to this problem is given by \cite{paper2}. In this paper, they propose a change in the hierarchy by changing the size of smaller caches so as to improve the average cache access latency. This paper uses the server workloads as the application instance for which the improvements are made. This is an important application instance which works on multi-core servers, but the cache hierarchies are not necessarily designed targeting server applications.

The modern processors are designed with three-level cache hierarchy having small L1 and L2 for fast cache access latency and a large shared LLC so as to accommodate for varying cache capacity demands. This type of hierarchy with small private L2 cache is a good design for those applications which fit into the L2 cache size. However, for bigger applications that are much larger than the L2 cache size, this design results in degraded performance as most of the execution time is spent on on-chip interconnect latency. While this can be solved by the design of a private or shared LLC and prefetching techniques to hide L2 miss latency, the access patterns for the server load applications are hard to predict and too complex for industry use. Therefore, the direct solution to reduce the overhead shared LLC latency by building large L2 caches is proposed in this paper as the better solution. The performance improvement for this methodology has been established in this paper by use of simulations of server workloads on a 16 core CMP.

By increasing the L2 cache size, more of the workload will be serviced at the L2 hit latency instead of the shared cache access latency. The simulations showed that by increasing the L2 cache size alone, without any change to the other properties, the performance improved by about 5\% both in the presence and the absence of prefetching techniques. Also, a sensitivity study was implemented and found that performance improvement was predominantly due to servicing code requests at L2 cache hit latency, suggesting the improvement of L2 cache management by preserving latency critical code lines over the data lines as a possible design change.

Though the performance seems to have improved, there are other drawbacks of this methodology that need to be corrected for a successful implementation. It was seen that though the average cache access was improved by increasing the L2 cache access latency for the server loads, the increase in size can significantly affect the performance of those with a working set that is fit into the smaller L2 cache. Furthermore, any change in the size of the caches needs to conform to the design properties of the cache hierarchy. For inclusive cache hierarchy, the L2:LLC area ratio needs to be maintained in order to support the inclusion property. Ignoring this would result in a waste of cache capacity and also create a negative effect of inclusion. On the other hand, increasing the LLC size to overcome this is not a good option. Hence, the size of the cache could be increased by either stealing the space from the LLC itself or going for exclusive or non-inclusive hierarchy instead.

The paper proposes an exclusive hierarchy that meets the design requirements and the manufacturing constraints. The L2 size can be increased while the constraints of the total on-chip area for cache space is maintained if inclusion is relaxed. It was seen that by moving to exclusive hierarchy, the effective caching capacity is maintained while the average cache access latency can be improved. Though, this comes at a cost, as increasing the L2 cache can reduce the observed shared LLC capacity which degrades the performance of the bigger workloads which might have fit in the base LLC but not of the smaller ones. Besides, with an exclusive hierarchy where duplication is not allowed, longer access latency might be produced in order to service data. Despite these trade-offs, the results showed a significant performance increase for the commercial workloads. 

To further better the performance of using exclusive hierarchy, the authors propose the use of including a single bit per L2 cache line called Serviced From LLC (SFL) bit for Re-Referenced Interval Prediction (RRIP) to improve the cache hit rates. An exclusive hierarchy invalidates lines on cache hits and this creates a new challenge for cache replacement policies which are meant to preserve lines that receive cache hits. However, in an exclusive LLC, the re-referenced cache lines are not preserved and using the same policy proposed for the inclusive hierarchy does not provide any improvement on the performance either. Thus, the author proposes storing the re-reference information in the L2 cache using SFL bits. This was illustrated using RRIP replacement policy. It was seen that the conventional RRIP actually reduced the hit-rate of 25\% while the use of SFL bit in the L2 cache helps replace the RRIP functionality and considerably improves the hit rates by 50\%.

The authors also propose preserving latency critical code lines in L2 caches over the data lines by using Code Line Preservation (CLIP). By implementing this, the majority of the L1 instruction cache misses are serviced by the L2 cache. Even though this allows preservation of the code lines in L2, if followed blindly, it can degrade the performance, especially when the working set does not contend for L2. Hence, CLIP is set to dynamically decide re-reference prediction of the data requests using Set Sampling. CLIP learns the code working set dynamically and the L2 capacity is allocated accordingly.

The results of these changes are studied in detail based on the simulation runs. It was seen that using SFL-bit in L2 caches to replace the RRIP functionality lost while migrating to exclusive hierarchy gives a performance boost of about 30\% in RRIP functionality with about 5\% performance degradation. It was also seen that by simple application of CLIP to the baseline cache, the performance of about half the server workloads increases by 5-10\%. This is equivalent to doubling the L2 cache size, which establishes the importance of code lines when compared to the data lines in the L2 cache. It was also seen that workloads with the highest front-end misses per 1000 instructions (MPKI) show more performance increase when L2 cache is increased along with CLIP. CLIP, though reduces front-end cache misses, increases the L2 cache misses. Still, the overall system performance is seen to increase significantly. 

Increasing L2 cache and decreasing LLC can have negative effects on workloads with working sets that fit in the larger shared LLC but not in the new LLC. This can have a performance decrease as high as 30\%. Although, for higher workloads such as the server loads, the larger L2 cache always showed better performance. With these improvements in place, the paper concludes showing about 5-12\% increase in performance of server workloads as opposed to the conventional hierarchy.

\subsection{Solution 3}

The third solution to the problem is the implementation of the Optical Caches in Chip Multiprocessor Architectures. The optical caches are being implemented on a separate chip rather than on the same CPU die. Spatial optical waveguides are used for the interconnection between the caches and CPU and the connection to the main memory. Wavelength Division Multiplexing (WDM) Optical interfaces are used for the cache interconnection systems. 

The main idea behind implementing the optical caches is to reduce the speed mismatch between the CPU and the main memory. The conventional ways of implementing a higher level of cache hierarchies not only increases the total area of the chip and the energy consumption but also the miss rates with more number of cores and higher cache size.

\begin{figure}[!h]
    \centering
    \includegraphics[width=0.35\textwidth]{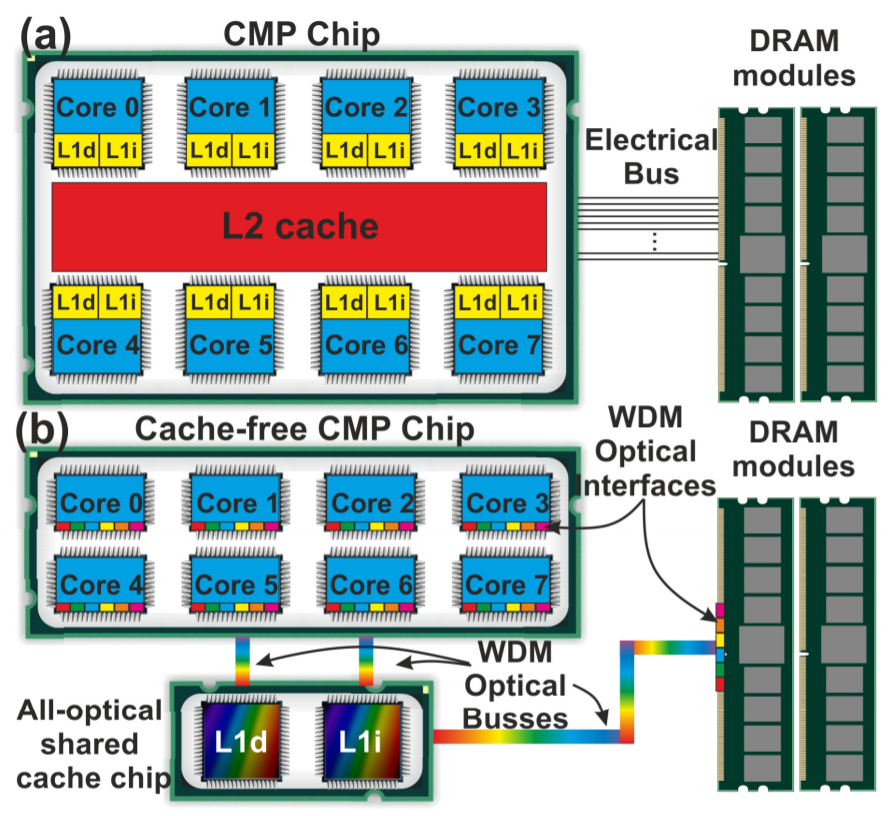}
    \caption{(a) Conventional Architecture (b) Shared Optical Cache Architecture ~\cite{paper3}}
    \label{fig : 3_arch}
\end{figure}

P. Maniotis et al.~\cite{paper3} proposed a solution to the above problem by presenting an optical cache memory architecture that uses optical CPU-MM buses, in place of standard electronic buses, for connecting all optical subsystems. In contrast to the standard way of putting the L1 and L2 cache along with the core in order to match their speeds, the optical caches are kept on a separate chip, as shown in fig. 1. A single-level, shared optical cache chip is used in the proposed architecture that lies next to the CMP chip and consists of separate L1-instruction (L1i) and L1-data (L1d) caches. That means that CPU chip is free from caches and therefore, the die area is reduced and can be used to add more cores to the CPU die. 

Additionally, WDM interfaces are used at the edges of the cores and the main memory for proper connection between the subsystems using optical buses.  The three optical buses consist of multiple communication layers, the number of which depends on their operations. The first layer is CPU-L1d, used by the CPU core to access the data from L1d. This again consists of 3 discrete layers, namely the address, data-write, and data-read. The second layer is CPU-L1i, which loads instructions from L1i to the CPU core. Unlike CPU-L1d, this layer only consists of two layers: address and data-read. The final layer is the L1-MM bus, which executes read-write operations between the cache chip and MM.

WDM is used to send various bits over a single waveguide, in contrast to the conventional ones in which n-bits are sent over n-parallel wire lanes. In the proposed architecture, a pair of wavelengths are used, one for sending the actual bit and another for sending its complement value. This means that transferring 8-bits worth of information would require 16 wavelengths per waveguide. An optical coupler is used for the communication between the cores and waveguides, which propagates the data power in the bus. However, a small amount of power is dropped at its attached core. For optical waveguides, Silicon-on-Insulator (SOI) technology or single mode polymer waveguides are used, as it causes low propagation losses of about 0.6 dB/cm and a small total density of 50 wires/cm. In optical interfaces, filters and modulators based on ring modulators,  operating at the speed of 10 Gbps, are used for optoelectronic and electro-optical conversion. The transmitter module has a ring modulator that is used to modulate CW signal and is tuned at a specific wavelength. For a 64-bit interface, a total of 8 transmitters are used, and each waveguide carries 8-bits of information. In the receiver module, add or drop ring resonator filters tuned to different wavelengths are used. A photo-detector module is used to convert any dropped wavelength into an electric signal which is then stored into a bit-register.

The optical caches consist of 5 subsystems: Read/Write Selector (RWS), Row Decoder (RD), Way Selector (WS), 2D Optical RAM Bank (2DRB), and Tag Comparator (TC). The RWS is used for read/write operations and for permitting and prohibiting the access to the 2DRB's content. The RD is used for activation of the 2DRB lines on the basis of address line fields requested by CPU. The WS is only used during writing operations to forward tag bit signals and the incoming data to the proper cache way. Data and tag parts of the cache are incorporated in the 2DRB and are implemented using all-optical flip-flops. They are grouped together under the same optical access gate after every 8 flip-flops using Arrayed Waveguide Grading (AWG) multiplexer. To determine the cache hit and the miss rate, a Tag Comparator is used. It consists of a set of all-optical XOR gates which compare the tag bits received from the CPU with the ones being emitted from both ways of the RD-activated 2DRB’s row.

From simulation results, it can be observed that a high-quality final cache output with an average extinction ratio of 12 dB and up to 16 GHz speed for both writing and reading error-free operations is obtained. In addition, Photonic Crystal (PhC) nanocavity technology provides compact multiplexer modules, switch and flip-flops that can be scaled to multi-bit storage devices and can result in completing optical cache memory solutions of low energy consumption in comparison to the electronic units. The CMP architecture of the off-chip cache module is being compared with the conventional L1-L2 hierarchy and the L1 (with no L2 cache) hierarchy. In contrast to the current L1 cache in CMPs, a single-level L1 optical cache is shared among all the cores. The optical caches memory proved to be very fast in serving multiple requests arriving from different cores into a single electronic core cycle without stalling the execution. However, in the case of shared electronic caches, L2 can be a bottleneck subject to multiple L1 cache misses. As such, the latency could be in the order of tens of cycles. Optical cache flat hierarchy also helps in solving the problem of data consistency that is contingent upon an electronic multi-level design, where the same data being cached in different units of cache gets updated. 

For simulations, a CMP is used with a number of cores varying from 2 to 16 with powers of 2, a core speed of 2GHz, an MM of speed 1GHz and size 512MB, an optical caches speed of 2xN, while the speed of buses Core-L1, L1-MM is same as the optical caches clock using optical waveguides. The simulations are performed on the 6 benchmarks of the PARSEC suite, which includes 13 multi-threaded unique programs covering diverse workloads for a variety of application domains. These benchmarks combine medium and high data exchange properties with either low or high data sharing properties.

\begin{figure}[!h]
    \centering
    \includegraphics[width=0.35\textwidth]{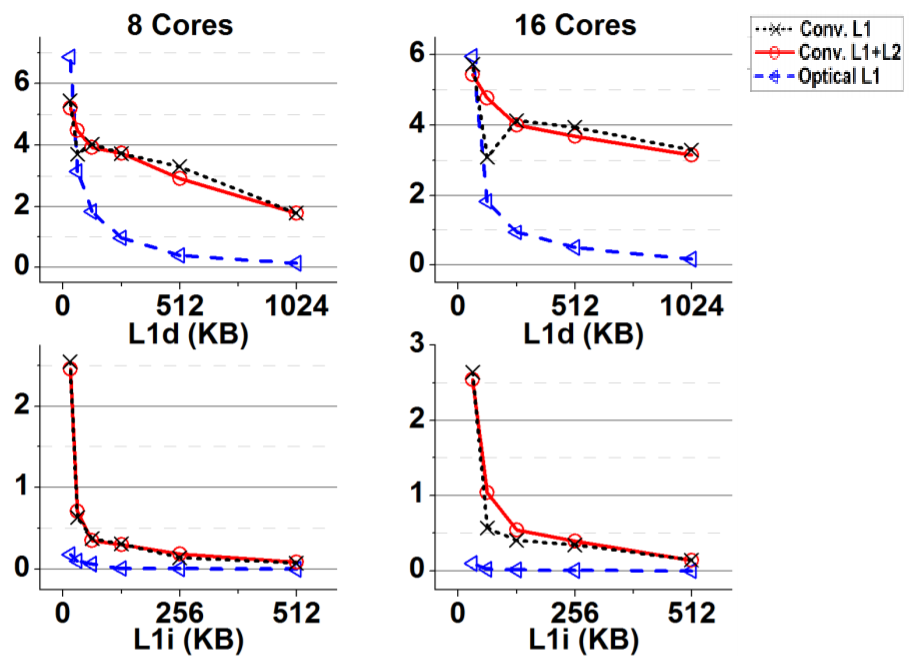}
    \caption{bodytrack’s miss rates of L1i and L1d caches for Conv. L1, Conv. L1+L2 and Optical L1 architecture for 8 and 16 cores~\cite{paper3}.}
    \label{fig : 3_miss_rate}
\end{figure}

The simulation results reveal that miss rates for conventional architectures are higher than the optical L1 architecture. Miss rates in conventional architecture increase as the number of cores increases, but not much variation is observed in other cases. L1i and L1d cache miss rates increase with the conventional architecture, but Optical L1 architecture is not affected much since efficient inter-thread data sharing and exchange is allowed by shared cache topology. Amplified performance of up to one-tenth of the conventional architecture is obtained with $N=8, 16$ and L1d size of 512KB and 1024KB, as can be analyzed from fig. 2. It is observed that all the programs in the PARSEC suite had similar results to the body track program, having L1d miss rates equal to a fraction of conventional architecture values. Optical L1 architecture handles L1i better than the conventional architecture even for small cache sizes. As far as execution cycles are concerned, optical caches provide better execution times than the conventional architecture for the same cache sizes, revealing the better performance of the former. Execution times reduce significantly as cache size is increased.

So, the paper concludes that optical shared cache architecture provides much better performance in comparison with both conventional architectures. Not only does it support high data exchange and high parallelization, but it also provides ultra-fast caching, an off-chip cache that helps in reducing the die area, and also a lower cache capacity. For certain cases, the miss rate reduction is up to 96\%, the average performance speed is increased up to 20.52\%, and the reduced average cache capacity is increased up to 65.8\%.

\section{Comparison}


The three solutions, though they all discuss various methods to increase the performance of high workload applications, have very different strategies that they use to achieve the expected results. For example, the Solutions 1 and 2 seem to use opposing strategies where one increases L2 cache to decrease on-chip interconnect latency while the other discusses decreasing L2 to the extent of completely eliminating L2 such that L1 serves most of the cache requests. Solution 3 agrees with Solution 1 in completely eliminating L2, but adds the use of optical caches to improve the performance and speed of operation. Nevertheless, all three methodologies contribute to significant improvements when working with high workloads.

\begin{figure}[!h]
    \centering
    \includegraphics[width=0.45\textwidth]{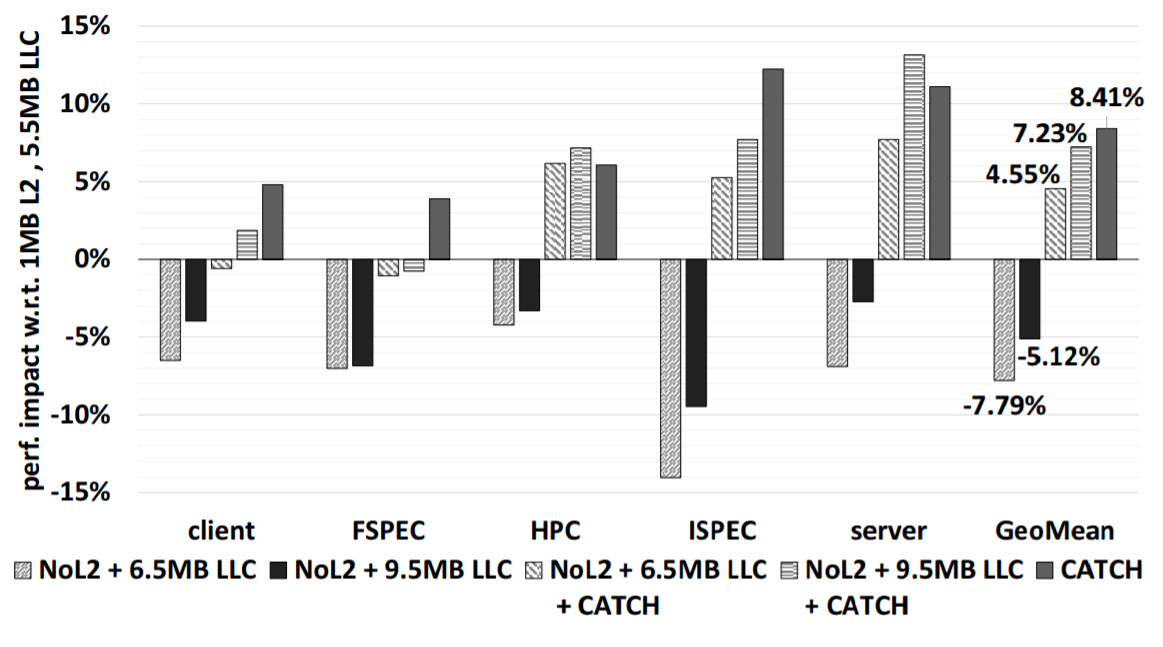}
    \caption{Performance graph for Solution 1~\cite{paper1}}
    \label{fig : perf_1}
\end{figure}

\begin{figure}[!h]
    \centering
    \includegraphics[width=0.45\textwidth]{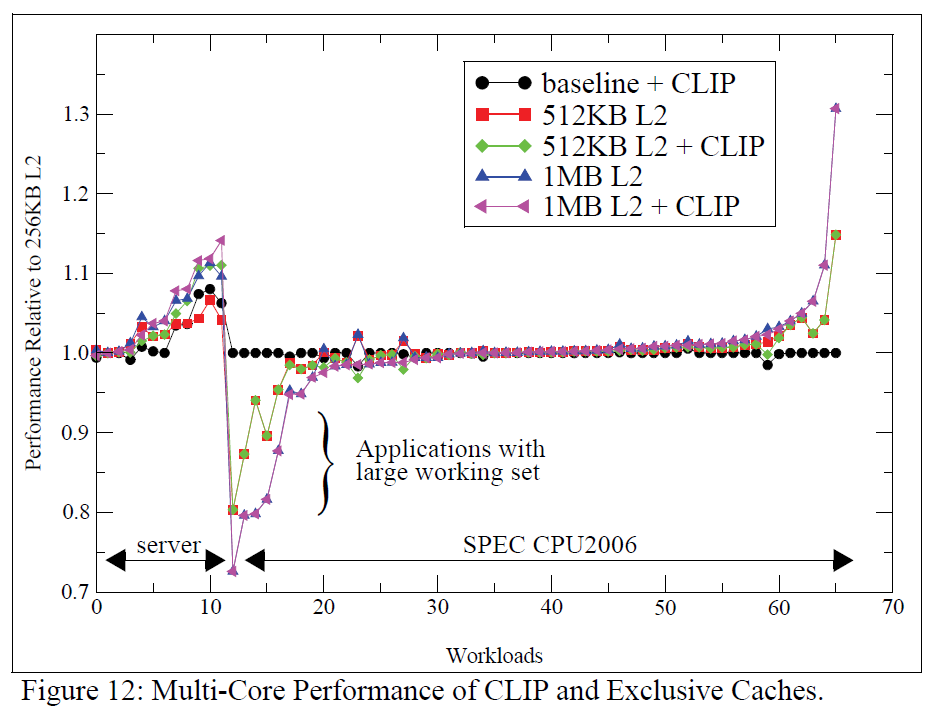}
    \caption{Performance graph for Solution 2~\cite{paper2}}
    \label{fig : perf_2}
\end{figure}

\begin{figure}[!h]
    \centering
    \includegraphics[width=0.2\textwidth]{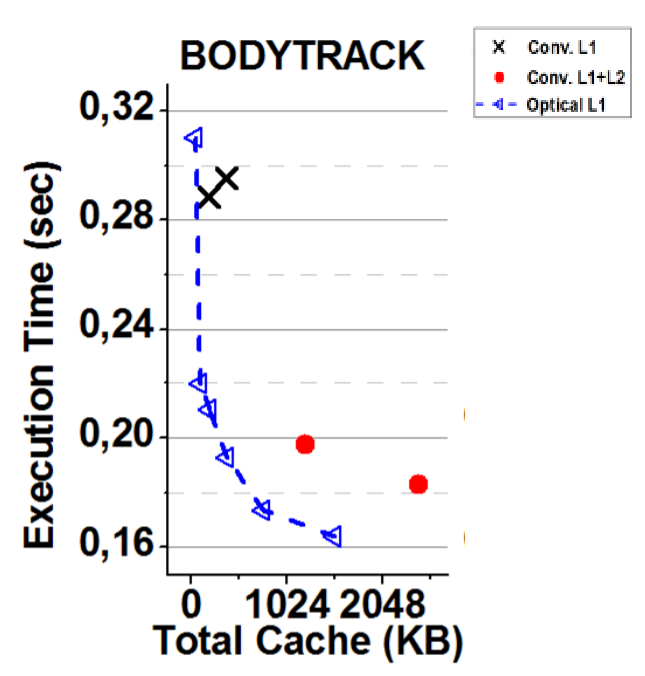}
    \caption{Performance graph for Solution 3~\cite{paper3}}
    \label{fig : perf_3}
\end{figure}

Figure \ref{fig : perf_1} shows the performance improvement of Solution 1. It can be seen from this figure that completely eliminating L2 while implementing CATCH gives a performance boost of up to 13\% for the server workloads. Similarly, Figure \ref{fig : perf_2} shows the performance increase of Solution 2, where it can be seen that the performance for server loads increase by 5 to 15\% with a 1MB L2 and CLIP. The performance of Solution 3 can be seen from Figure \ref{fig : perf_3} where the execution cycle and consequentially the performance improves by about 20\%.

Even though all these models show a high boost in performance, they have their own advantages and disadvantages. Apart from the gain in performance, Solution 1 has a major advantage of a smaller area, which can be exploited to increase the cores of the processor. Moreover, the CATCH framework requires a considerably small size of about 4 KB in hardware. The paper, unfortunately, fails to provide a solution to the high interconnect latency, a problem of which is well addressed in Solution 2. In stark contrast, Solution 2 discusses expanding the L2 cache to enhance the performance and simultaneously reduce the interconnect latency. Although, as a consequence, it comes with a disadvantage in terms of area consumption and thus a fewer number of core processors. Furthermore, this solution creates a negative performance profile for other application instances where the workload is much smaller which just as well could fit in the conventional L2 cache sizes or for bigger workloads which fit in the conventional LLC space. 

Solution 3 addresses both these problems of area consumption and interconnect latency by providing a new solution of also removing the L2 cache, but combined with the use of optical shared caches. As discussed earlier, removing L2 cache decreases the on-die area which creates space for the inclusion of more cores, while the use of optical caches provides ultra-fast caching. This solution maintains the performance of conventional architectures for smaller workloads and at the same time provides higher speed and better performance for bigger workloads.

\balance
\section{Conclusions}
In this report, we have discussed three different cache hierarchy designs that could improve the performance of systems with high workloads. In Solution 1, we described a new inter-cache prefetching technique to process instructions according to their criticality, while also arguing about the possible benefits of reducing or possibly removing the L2 cache. This solution helps in improving the performance by up to 13\% for server workloads. In Solution 2, we discussed increasing the size of L2 instead in order to improve average cache access latency. Not only does this help in reducing the on-chip interconnect latency, but it also elevates the performance for server loads from 5 up to 15\%. In Solution 3, we considered the implementation of off-chip shared optical caches, which are connected to the CPU core and the MM through optical waveguides. This results in better area utilization that can then be used to either instill more cores or simply reduce the die area. 

WDM optical interfaces, along with the spatial-multiplexed optical waveguides, are used to connect the cores-cache units and the MM-cache units. Such communication takes place in the optical domain. This architecture helps not only in significantly reducing the L1 miss rate up to 96\% in some cases but also in improving the performance by 20.52\% along with reducing the average cache capacity up to 65.8\%. The architecture proved to be useful in the case of high parallelization and heavy server loads. Ultimately, we conclude that all the above-discussed solutions are able to significantly boost the performance for heavy server workloads, although Solution 3 comes up as the best solution in terms of its benefits and drawbacks.

\begin{appendices}

\end{appendices}
\end{document}